\documentclass[sigconf]{acmart}
\usepackage{booktabs} 
\usepackage{caption,subcaption} 
\renewcommand\footnotetextcopyrightpermission[1]{} 

\newcommand{\task}{\mathcal{T}}

\begin{document}

\title{Discriminatory Transfer}

\author{Chao Lan}
\affiliation{%
  \institution{University of Kansas}
  \streetaddress{2335 Irving Hill Rd.}
  \city{Lawrence} 
  \state{Kansas} 
  \postcode{66045}
}
\email{pete.chaolan@gmail.com}

\author{Jun Huan}
\affiliation{%
  \institution{University of Kansas}
  \streetaddress{2335 Irving Hill Rd.}
  \city{Lawrence} 
  \state{Kansas} 
  \postcode{66045}
}
\email{jhuan@ittc.ku.edu}

\begin{abstract}
We observe standard transfer learning can 
improve prediction accuracies of target 
tasks at the cost of lowering their prediction 
fairness -- a phenomenon we named 
\textit{discriminatory transfer}. 
We examine prediction fairness of a standard 
hypothesis transfer algorithm and a standard 
multi-task learning algorithm, and show they 
both suffer discriminatory transfer on the 
real-world Communities and Crime data set. 
The presented case study introduces an 
interaction between fairness and transfer 
learning, as an extension of existing fairness 
studies that focus on single task learning. 
\end{abstract} 

\maketitle

% It is observed a standard predictive learning task 
% can admit a model of high prediction accuracy 
% but low prediction fairness, i.e. its predictions 
% are illegally biased against disadvantaged instances. 
% This phenomenon is commonly referred as 
% \textit{fairness-accuracy trade-off}. Thus, how to 
% improve fairness without over-compromising original 
% accuracy is an ongoing research direction. 

% However, existing studies focus on a single-task setting, 
% while in many problems one can learn \textit{multiple} 
% predictive tasks together in hope of boosting their accuracies 
% by transferring information among them -- a topic generally 
% referred as transfer learning \cite{pan2010survey}. 

% suggests a new cause of illegal bias 
% in model predictions -- the process of information transfer. 
% It suggests transfer learning researchers start monitoring 
% fairness of their target tasks, and considering compromising 
% the traditional accuracy boost for maintaining prediction 
% fairness of these tasks. 

\section{Introduction}

It has been observed a predictive learning task can 
admit a hypothesis of high prediction accuracy but 
low prediction fairness (i.e. the model predictions 
are illegally biased against disadvantaged instances). 
This has raised an argument for the fairness-accuracy 
trade-off in machine learning, and many studies propose 
methods to balance prediction fairness and accuracy. 

However, existing studies focus on a single learning task, 
while in many problems one can (formalize and) jointly learn 
multiple related tasks in hope of boosting their overall 
prediction accuracy -- a topic commonly referred 
as \textit{transfer learning} \cite{pan2010survey}. 
\textit{What is the interplay between fairness and transfer 
learning}? This motivates the presented study. 
\let\thefootnote\relax\footnote{This paper is presented as a 
poster at the 2017 Workshop on Fairness, Accountability and 
Transparency in Machine Learning (FAT/ML 2017).}

In this paper, we show standard transfer learning can improve 
prediction accuracies of participating tasks (as usual), 
but at the price of lowering their \textit{originally high} 
prediction fairness. We name this phenomenon 
\textit{discriminatory transfer}. 
Our study reveals the information transfer process as a new 
cause of unfair model predictions, even when each participating 
task by itself admits a model with fair predictions. 
How does discriminatory transfer happen, and how to mitigate 
it while retaining the traditional gain of information transfer 
as much as possible remain open questions. 

Our empirical examination is based on two popular and generic 
transfer learning algorithms, i.e.  
the hypothesis transfer algorithm formalized by Kuzborskij and 
Orabona\cite{kuzborskij2013stability} and the multi-task algorithm 
formalized by Ciliberto et al \cite{ciliberto2015convex}. 
The two algorithms correspond to two settings: the former assumes 
one task is pre-learned and its solution is used to assist learning 
another task; the latter assumes all tasks are learned jointly with 
constraints on task relatedness. 
Our main fairness measure is based on equalized odds proposed by 
Hardt et al \cite{hardt2016equality}, which is a conditional refinement 
of the legal notion \textit{disparate impact}. We experiment  
on the Communities and Crime data set \cite{redmond2002data}.

\section{Preliminaries}

\subsection{Notations and Problem Setting} 

Without loss of generality, we consider two supervised 
learning tasks $\task$ and $\task'$. 
Task $\task = ( X, P, f_{*} )$ consists of 
a population $X$, a distribution $P$ on $X$, 
and a target predictive function $f_{*}: X \rightarrow [0, 1]$; 
the goal is to learn $f_{*}$ based on a random sample $S$ 
drawn from $P$ and labeled by $f_{*}$, 
plus any information transferred from the other task. 
Task $\task' = ( X', P', f'_{*} )$ is defined in 
a similar manner. 
We assume both $X, X'$ are embedded in $\mathbb{R}^{p}$ 
for some $p>0$, but they are not necessarily identical. 

\subsection{Hypothesis Transfer Algorithm} 

Hypothesis transfer is a popular transfer learning paradigm, 
which aims to improve prediction accuracy of a task by 
additionally using a pre-learned predictive model of another 
related task. For more information on this topic, 
see \cite{kuzborskij2013stability} and the reference therein. 

We will examine a generic algorithm called Regularized Least 
Square (RLS) formalized and theoretically justified 
by Kuzborskij and Orabona \cite{kuzborskij2013stability}. 
The algorithm focuses on linear predictive functions, 
i.e. any function $f(x) = x^{T} w$ is uniquely parameterized 
by a $p$-dimensional vector $w$. 
Suppose task $\task'$ is pre-learned and its learned 
predictive function $\hat{w}'$ is used to assist task $\task$. 
Let $x \in S$ be an training example of task $\task$ 
and $y = f_{*}(x)$ be its label. The RLS algorithm 
solves the following problem 
\begin{equation}
\label{hypotransferproblem}
\min_{w} \frac{1}{|S|} \sum_{x \in S} (x^{T} w - y)^{2} 
+ \lambda \parallel w - \hat{w}' \parallel^{2},   
\end{equation}
where $\lambda$ is a regularization coefficient. 
The authors showed the solution $\hat{w}$ to 
(\ref{hypotransferproblem}) has an analytic form  
\begin{equation}
\hat{w} = X ( X^{T} X + |S| \lambda \bf{I})^{-1} 
(Y - X^{T} \hat{w}') + \hat{w}',  
\end{equation}
where $X \in \mathbb{R}^{|S| \times p}$ is a data matrix 
with row $i$ representing example $i$, 
and $Y \in \mathbb{R}^{|S|}$ is a label vector with 
element $i$ representing the label of example $i$. 

It is clear $\lambda$ controls the degree of information 
transferred from task $\task'$ to task $\task$, 
in a sense that larger $\lambda$ will bias $\hat{w}$ 
towards $\hat{w}'$ more strongly. 

When two tasks do have similar target hypotheses, 
one may expect reasonably larger $\lambda$ leads to a 
more accurate $\hat{w}$. (Of course, improperly 
large $\lambda$ may mis-bias learning and lower the 
accurate, a problem known as negative transfer.) 
While this used to be a happy ending, 
our experimental study will show larger $\lambda$ 
can meanwhile hurt prediction fairness.

\subsection{Multi-Task Algorithm}

Multi-task learning is another popular transfer learning 
paradigm, which aims to improve prediction accuracies of 
both tasks by learning them \textit{jointly} with proper 
constraints on task relations. 

We will examine a recent multi-task learning algorithm 
developed by Ciliberto et al 
at \footnote{https://github.com/cciliber/matMTL}. 
Recall $S$ is a random sample of task $\task$ with 
example $x$ and its label $y$. Similarly, let $S'$ 
be a random sample of task $\task'$ with example $x'$ 
and its label $y'=f_{*}'(x)$.  
The algorithm finds $f, f'$ that minimize the following problem 
\begin{equation}
\label{mtlalgorithm}
\frac{1}{|S|}\sum_{x \in S} \mathcal{L}(y, f(x)) 
+ \frac{1}{|S'|}\sum_{x' \in S'} \mathcal{L}(y', f(x')) 
+ \lambda ||\vec{f}||_{\mathcal{H}}^{2},   
\end{equation}
where $\mathcal{L}$ is loss function and 
$\mathcal{H}$ is Hilbert space of vector-valued 
functions $\vec{f}$ with scalar components $f, f'$.  
The term $||\vec{f}||_{\mathcal{H}}^{2}$ encodes 
relation between $f$ and $f'$ through a matrix $A$ 
such that, based on the Representer Theorem and 
other mild conditions, one has 
\begin{equation}
||\vec{f}||_{\mathcal{H}}^{2} 
= \sum_{t_{1}, t_{2}=1,2} A^{-1}_{t_{1}t_{2}} 
\cdot \left\langle f_{t_{1}}, f_{t_{2}}\right\rangle_{\mathcal{H}}, 
\end{equation}
where $f_{1}$=$f$ and $f_{2}$=$f'$. Micchelli and Pontil \cite{micchelli2005kernels} show proper choices of $A$ allow $||\vec{f}||_{\mathcal{H}}^{2}$ to capture certain 
established task relations.

Again, one may expect larger $\lambda$ to give more accurate 
$f$ and $f'$, provided they are similar. 
Our experimental study will show this is indeed the case, 
but increasing $\lambda$ also lowers prediction 
fairness of \textit{both} $f$ and $f'$. 

\subsection{Fairness Measure}

We will measure fairness based on the notion of equalized odds 
recently proposed by Hardt et al \cite{hardt2016equality}. 
In each task, suppose the population is divided into two 
groups $G_{1}$ and $G_{2}$, and one aims to examine prediction 
fairness between these groups. Let $f$ be a model for 
the task, we define the Equalized Odds Ratio (E.O.R.) of $f$ as  
\begin{equation}
\label{eor}
\text{E.O.R.} (f) = \frac{\text{Pr} \{ f(x) = 1 \mid x 
\in G_{1}, f_{*}(x) = 1 \}}{\text{Pr} \{ f(x) = 1 \mid x 
\in G_{2}, f_{*}(x) = 1 \}}.  
\end{equation}
Our proposed measure is first inspired by the well-known 
$80\%$-rule, which states $f$ gives fair prediction if the 
following ratio\footnote{We will refer this ratio 
as Disparate Impact Ratio (D.I.R.).} is no less than 80\% 
\begin{equation}
\label{ratio}
\frac{\text{Pr} \{ f(x) = 1 \mid x 
\in G_{1} \}}{\text{Pr} \{ f(x) = 1 \mid x \in G_{2} \}}.  
\end{equation}
However, it is argued the probabilities in (\ref{ratio}) do not 
guarantee equality of opportunity \cite{dwork2012fairness}, 
and authors in \cite{hardt2016equality} refine each probability 
by furthering conditioning it on $f_{*}(x)=1$ 
(i.e. not disparate impact \textit{within} truly qualified 
population and truly unqualified population).  
Our proposed measure is obtained by simply replacing 
the probabilities in (\ref{ratio}) with their refinements 
in \cite{hardt2016equality}. 

Similar to prior studies, we say the prediction fairness of $f$ is 
improved as $E.O.R.(f)$ approaches value 1, and is lowered as 
$E.O.R.(f)$ deviates from 1.

\section{Experiment} 

\subsection{Experimental Setup} 

We experimented on the Communities and Crime data set \cite{redmond2002data} 
obtained from the UCI data repository. 
The original data set collects for 1994 communities their crime rates 
and 122 normalized predictive attributes. The general goal of learning 
is to predict community crime rates based on these attributes.

We considered crime rate as label and binarized it such that crime rates 
above 50\% were encoded as 1 and others encoded as -1;  
the attribute ``percentage of population that is African American'' 
was considered sensitive and binarized such that percentages above 50\% 
were encoded as 1 and others encoded as 0 -- based on this, all 
communities were partitioned into either African American (AA) 
communities (percentages >50\%) or non-African American communities. 
We examined prediction discrimination against AA communities. 

We simulated four transfer learning settings based 
on the data set, each consisting of two tasks. 

\textbf{Setting 1} partitioned communities based on the 
\textit{population} attribute, such that task $\task$ is to 
predict crime rates for communities with populations greater 
than 0.05, and task $\task'$ for other communities\footnote{For 
all settings, our choice of the threshold (here, 0.05) was close 
to attribute mean, but also for demonstrating smoother results.}. 

\textbf{Setting 2} partitioned communities based on the 
\textit{median house income} attribute, such that task $\task$ 
is to predict for communities with incomes greater than 0.25, 
and task $\task'$ for other communities. 

\textbf{Setting 3} partitioned communities based on the 
\textit{number of people under poverty level} attribute, 
such that task $\task$ is to predict for communities 
with poverty count greater than 0.02, and task $\task'$ 
for other communities. 

\textbf{Setting 4} partitioned communities based on 
the \textit{number of homeless people in shelters} attribute, 
such that task $\task$ is to predict for communities with homeless 
count greater than 0.03, and task $\task'$ for the rest. 

In all settings, task $\task'$ was assumed pre-learned and 
used to assist task $\task$. When switching tasks, we observed 
similar trends but the results were numerically unstable and 
thus not reported here. 
Finally, all performance were evaluated over 10-fold cross 
validation, with 1 fold used for training and 9 for testing, 
and the averaged results were reported\footnote{We chose 1 fold 
for training to better demonstrate the well-known advantage of 
transfer learning on smaller samples}. 
Our source code is  at \url{https://github.com/petelan/FATML2017}.

\subsection{Example Results}

\begin{figure}[t!] 
\centering 
\begin{subfigure}{0.23\textwidth}
\centering
\includegraphics[width=\textwidth]{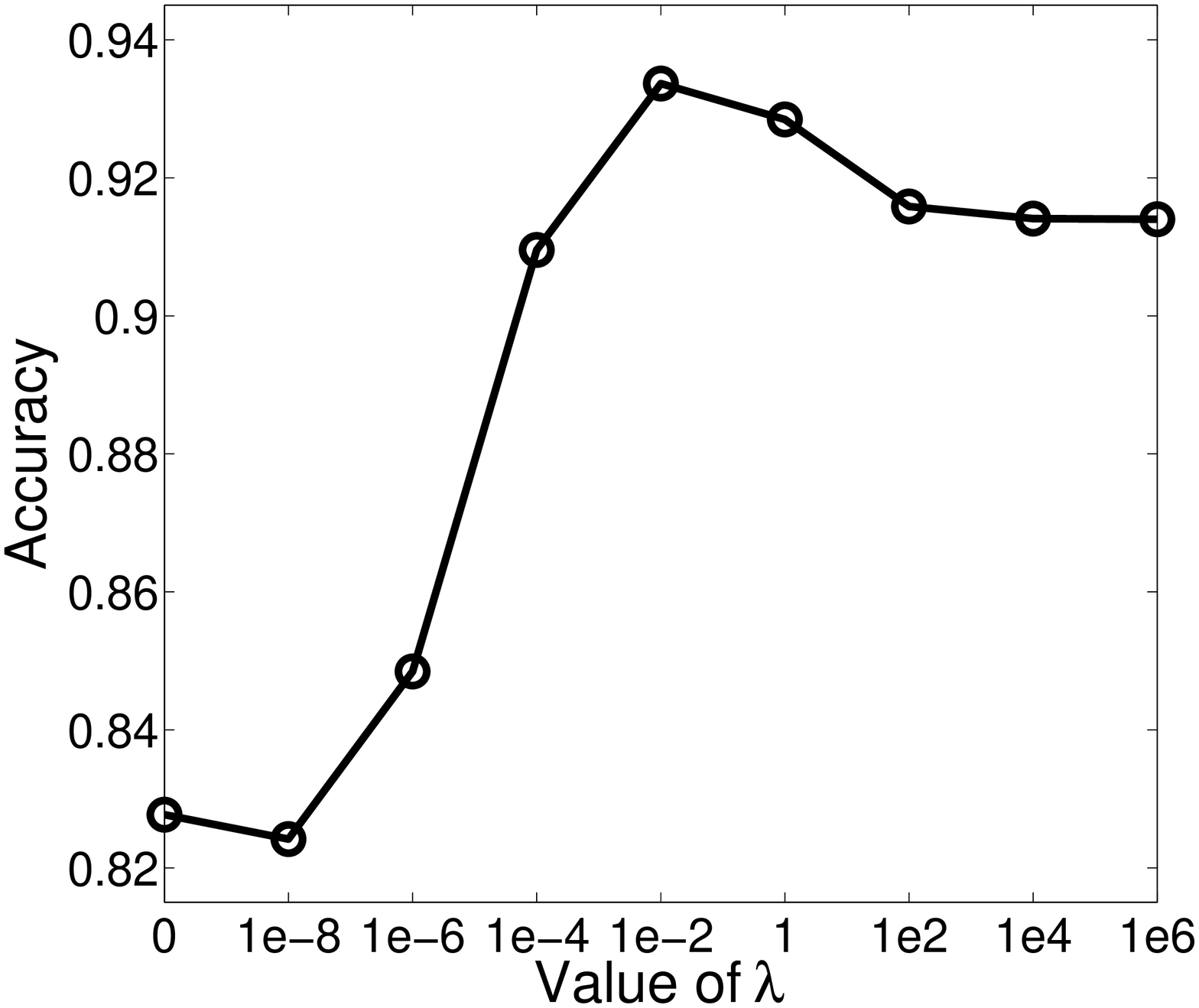}
\vspace{-15pt}
\caption{Accuracy} 
\end{subfigure}
\begin{subfigure}{0.23\textwidth}
\centering
\includegraphics[width=\textwidth]{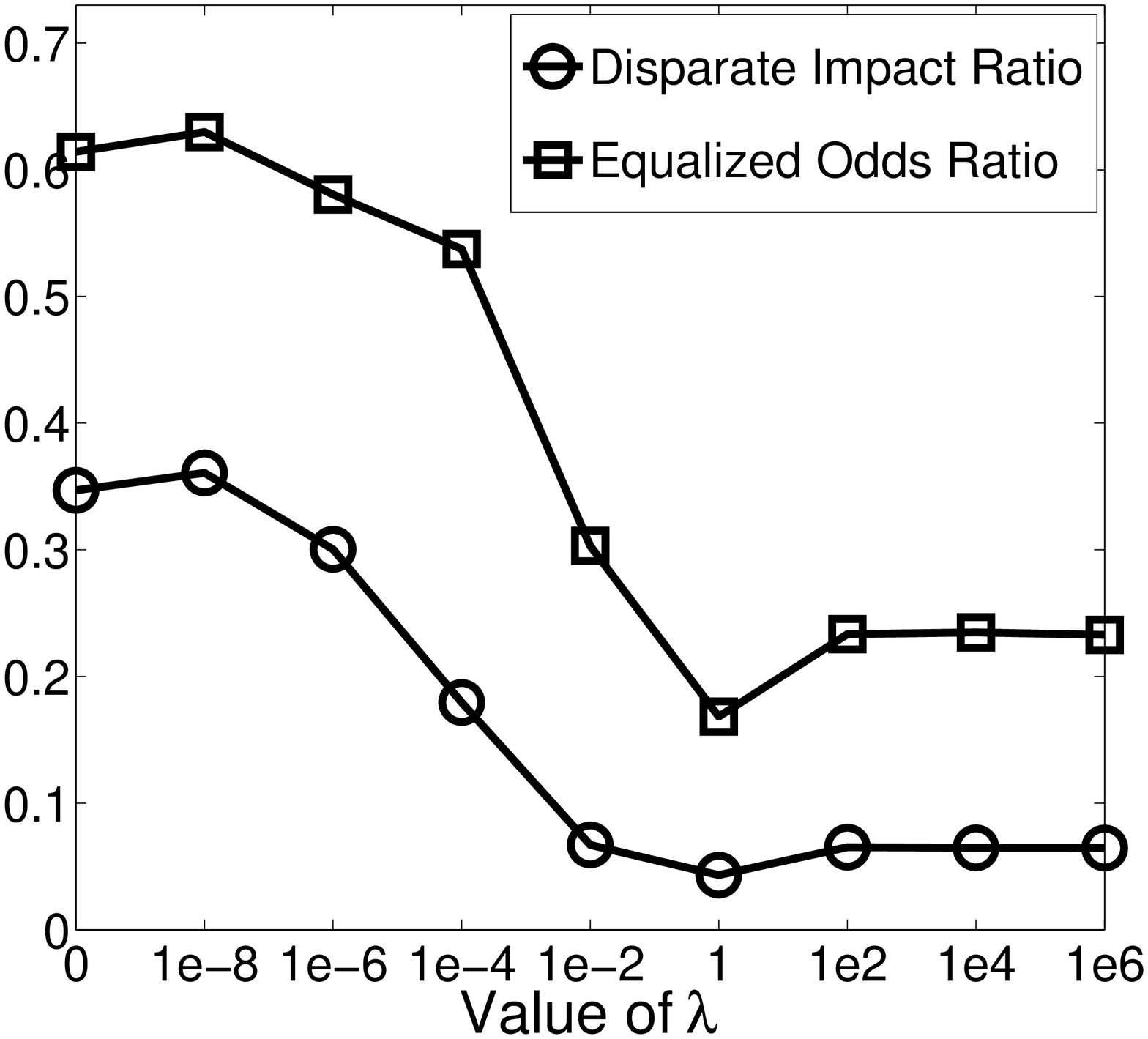}
\vspace{-15pt}
\caption{Fairness}
\end{subfigure}
\vspace{-10pt}
\caption{H.T. Performance under Setting 2. } 
\label{hyposet1}
\end{figure}

In this section, we show example results of discriminatory 
transfer in the hypothesis transfer (H.T.) algorithm under setting 2.
For the pre-learned task, we used ridge regression as the base 
learner, with regularization coefficient fixed to 1. 
To see the interaction between fairness and 
transfer learning, we varied coefficient $\lambda$ 
in (\ref{hypotransferproblem}) and obtained results 
in Figure \ref{hyposet1}. 

From Figure \ref{hyposet1} (a), we observe prediction 
accuracy is improved as $\lambda$ increases from $0$ to 
$0.01$, then mildly degrades and eventually converges. 
This is a somewhat common observation: it shows the traditional 
benefit of hypothesis transfer within range $[0,0.01]$, 
and implies the true task relation is captured at around 
$\lambda = 0.01$; as one reinforces task relation beyond 
that range, however, prediction accuracy can decrease. 

Our new observations come from Figure  \ref{hyposet1} (b).   
We see as $\lambda$ increases from 1e-8 to 1e-2, prediction 
accuracy is increased while prediction fairness is decreased 
-- this is the discriminatory transfer phenomenon. 

One may wonder whether the prediction fairness of the 
pre-learned hypothesis would have any impact on the result. 
Our next result suggests it may have an impact, 
but discriminatory transfer may still exist. 

\begin{figure}[t!] 
\centering 
\includegraphics[width=.5\textwidth]{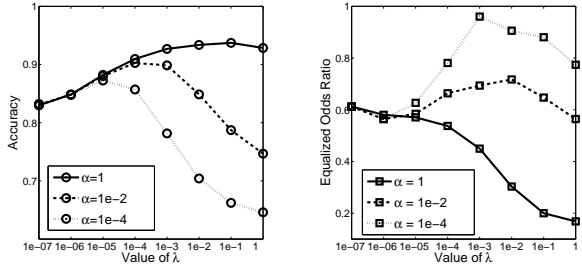}
\caption{H.T. Performance with different $\alpha$.}  
\label{figcmp}
\end{figure}

Recall we have fixed $\alpha = 1$ in the pre-learned task. 
We noticed decreasing $\alpha$ could result in hypotheses with 
higher prediction fairness though lower prediction fairness. 
See Table \ref{tab1}. 
So one may say our results in Figure \ref{hyposet1} is based on 
an `unfair' pre-learned hypothesis. 
We now repeat the same experiment but with smaller $\alpha$'s; 
results are reported in Figure \ref{figcmp}. 

Let us focus on the case when $\alpha$=1e-4, which means the 
pre-learned hypothesis is fair. 
As $\lambda$ increases from 1e-7 to 1e-6, we observe what may still 
be considered as discriminatory transfer -- prediction accuracy increases 
while prediction fairness decreases. As $\lambda$ continues to increase 
from 1e-6 to 1e-5, however, prediction fairness 
starts to grow back while prediction accuracy continues to increase. 
This is something new, and suggests there may be ways to mitigate 
discriminatory transfer. When $\lambda$ grows larger than 1e-5, 
prediction accuracy starts to drop (hence the traditional benefit of 
transfer learning is gone); 
we are not particularly interested in the results. 

How does discriminatory transfer occur? How to mitigate it while 
maximally maintaining the traditional accuracy improvement? 
What are other interactions between fairness and transfer learning? 
These are open questions. 
We conjecture standard transfer process may 
rule out some fair hypotheses when biasing target task learning.

\def\arraystretch{1.3}%
\begin{table}[t!]
\begin{tabular}{ccc}
\bf $\alpha$ & \bf Accuracy & \bf Fairness (E.O.R / D.I.R.)  \\ \hline 
1 & 0.7962 & 0.2908 / 0.1048 \\ 
1e-1 & 0.7518 & 0.4585 / 0.2532 \\ 
1e-2 & 0.6827 & 0.7125 / 0.4955 \\ 
1e-3 & 0.6277 & 0.9370 / 0.7152 \\ 
1e-4 & 0.6068 & 0.9787 / 0.7991 \\ 
1e-5 & 0.6039 & 0.9729 / 0.8105 \\ \hline 
\end{tabular}
\caption{Performance of Pre-learned Hypothesis in the Pre-learned Task.}
\vspace{-5pt} 
\label{tab1}
\end{table}

\subsection{Other Results} 

In this section, we show further evidence of 
discriminatory transfer on both hypothesis 
transfer and multi-task algorithms over four 
transfer learning settings. 
For convenience, in one experiment both prediction 
accuracy and fairness will be shown in the same figure; 
in these cases, the left vertical axis of the figure 
will represent accuracy and the right vertical axis will 
represent fairness (as measured by the equalized odds ratio).

\begin{figure}[t!] 
\centering 
\begin{subfigure}{0.23\textwidth}
\centering
\includegraphics[width=\textwidth]{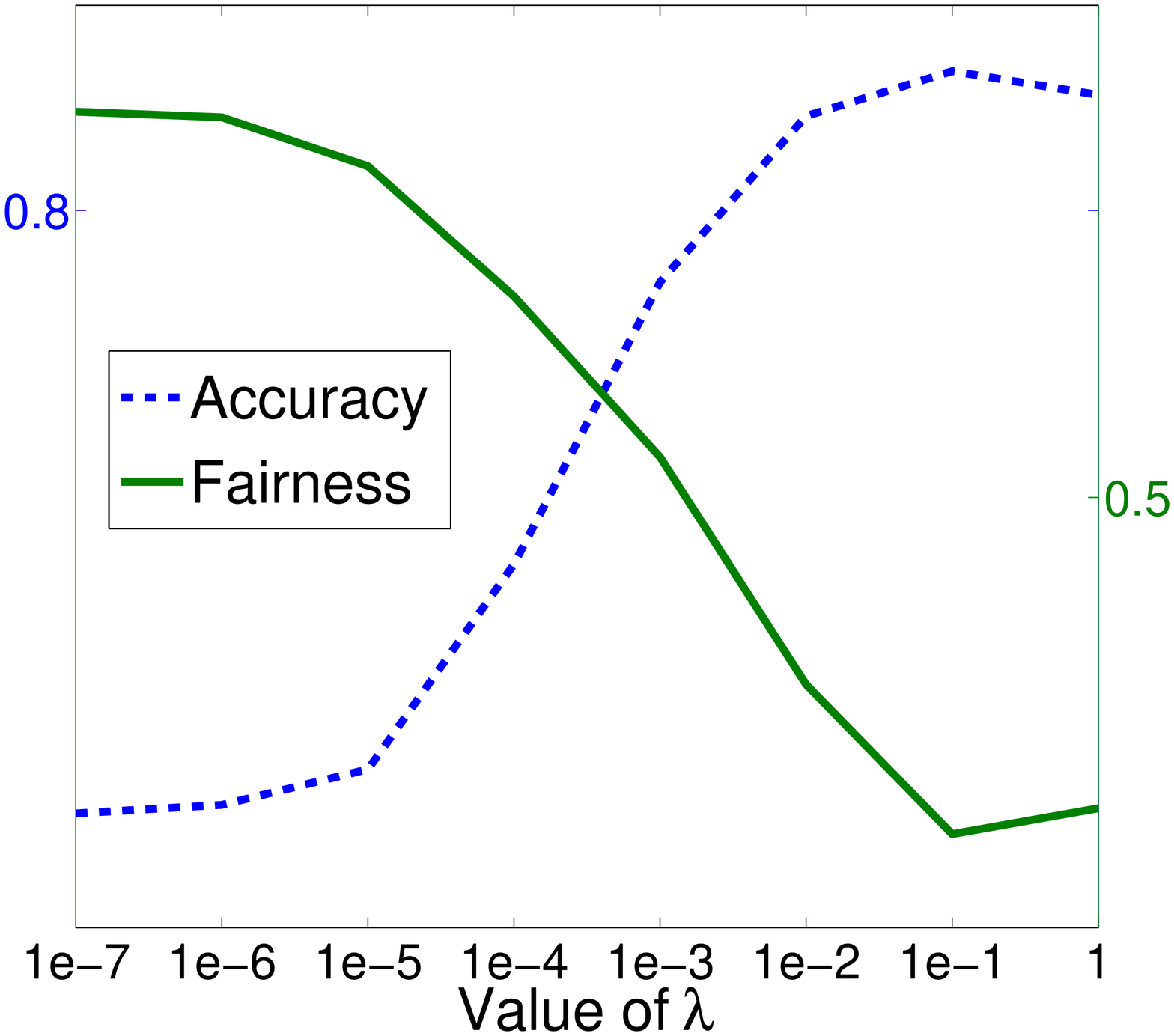}
\vspace{-15pt}
\caption{Setting 1} 
\end{subfigure}
\begin{subfigure}{0.23\textwidth}
\centering
\includegraphics[width=\textwidth]{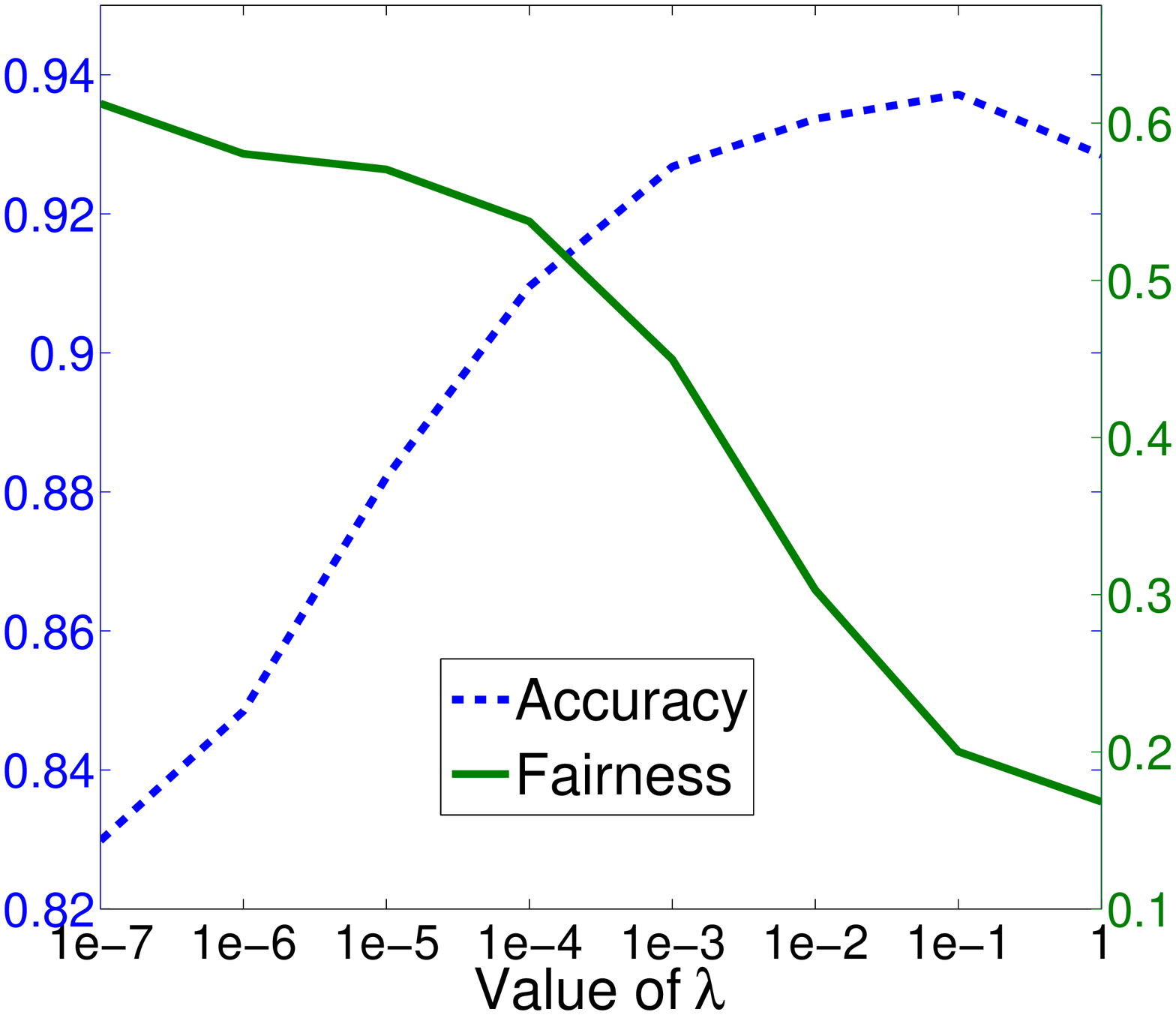}
\vspace{-15pt}
\caption{Setting 2}
\end{subfigure}
\begin{subfigure}{0.23\textwidth}
\centering
\includegraphics[width=\textwidth]{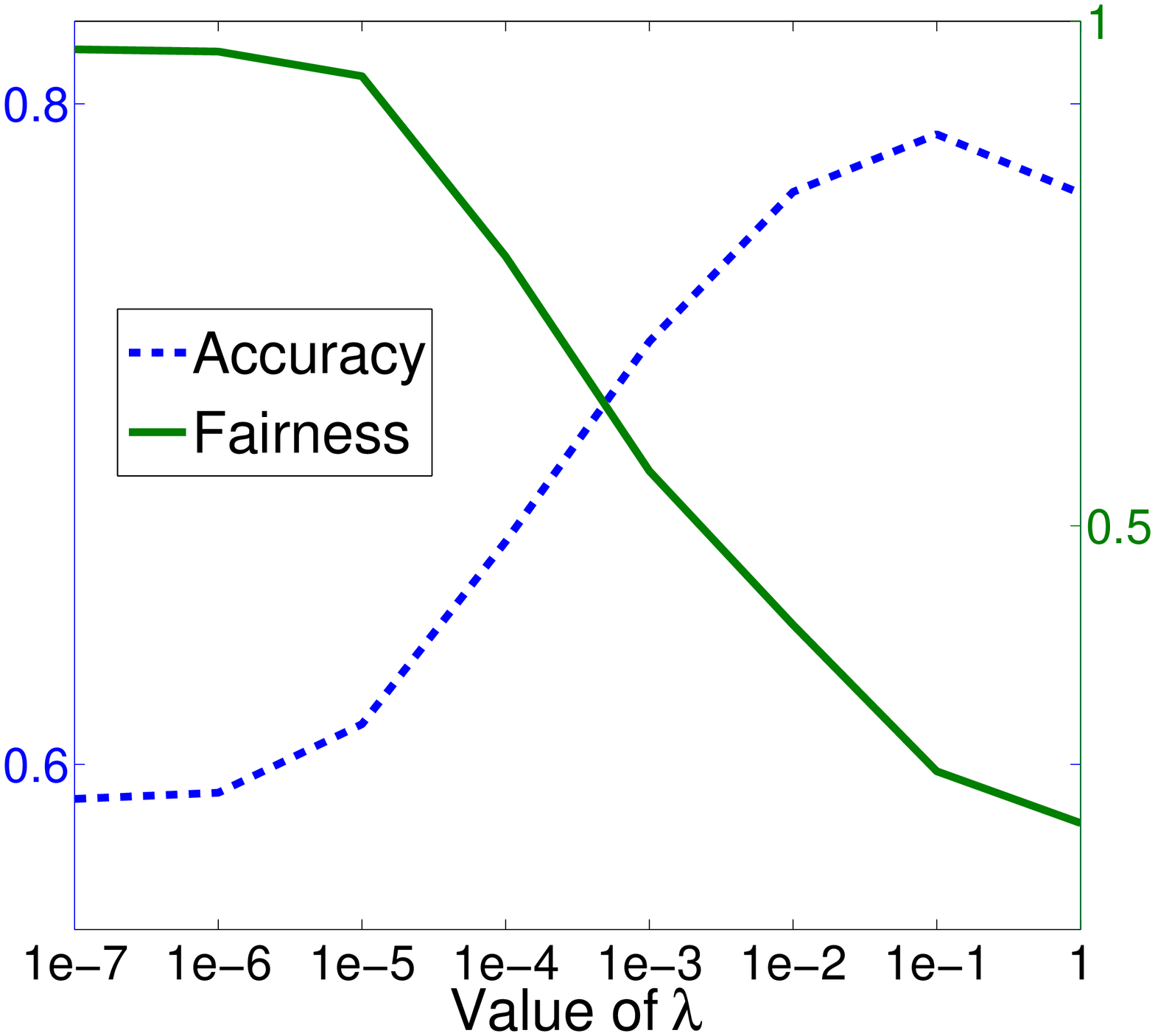}
\vspace{-15pt}
\caption{Setting 3}
\end{subfigure}
\begin{subfigure}{0.23\textwidth}
\centering
\includegraphics[width=\textwidth]{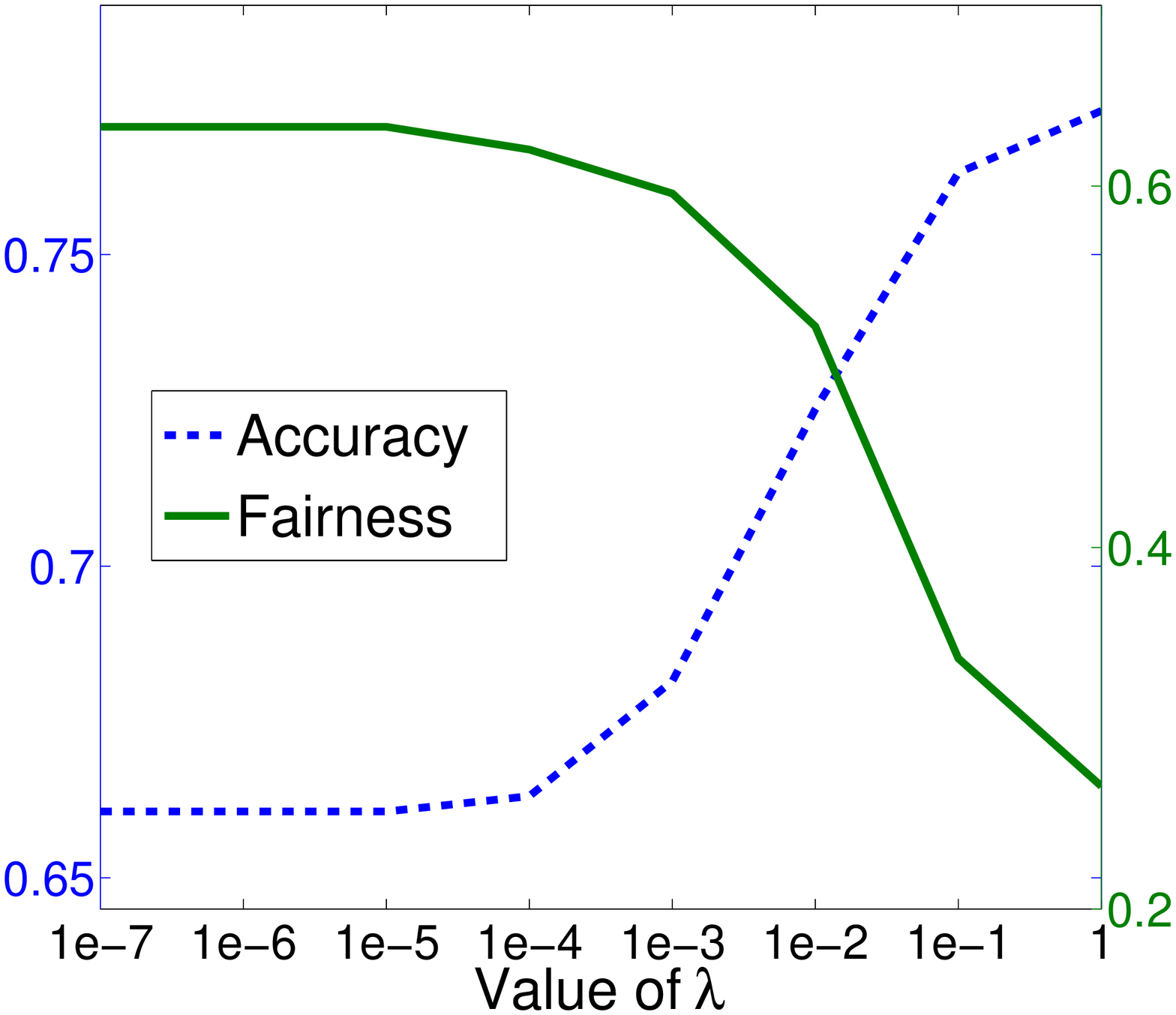}
\vspace{-15pt}
\caption{Setting 4}
\end{subfigure}
\caption{H.T. Performance in Four Settings. } 
\label{hypoallset}
\end{figure}

We first examined the performance of hypothesis 
transfer algorithm on four transfer settings. 
The results are shown in Figure \ref{hypoallset}. 
One can easily observe discriminatory transfer 
in all settings. 

\begin{figure}[h] 
\centering 
\begin{subfigure}{0.23\textwidth}
\includegraphics[width=\textwidth]{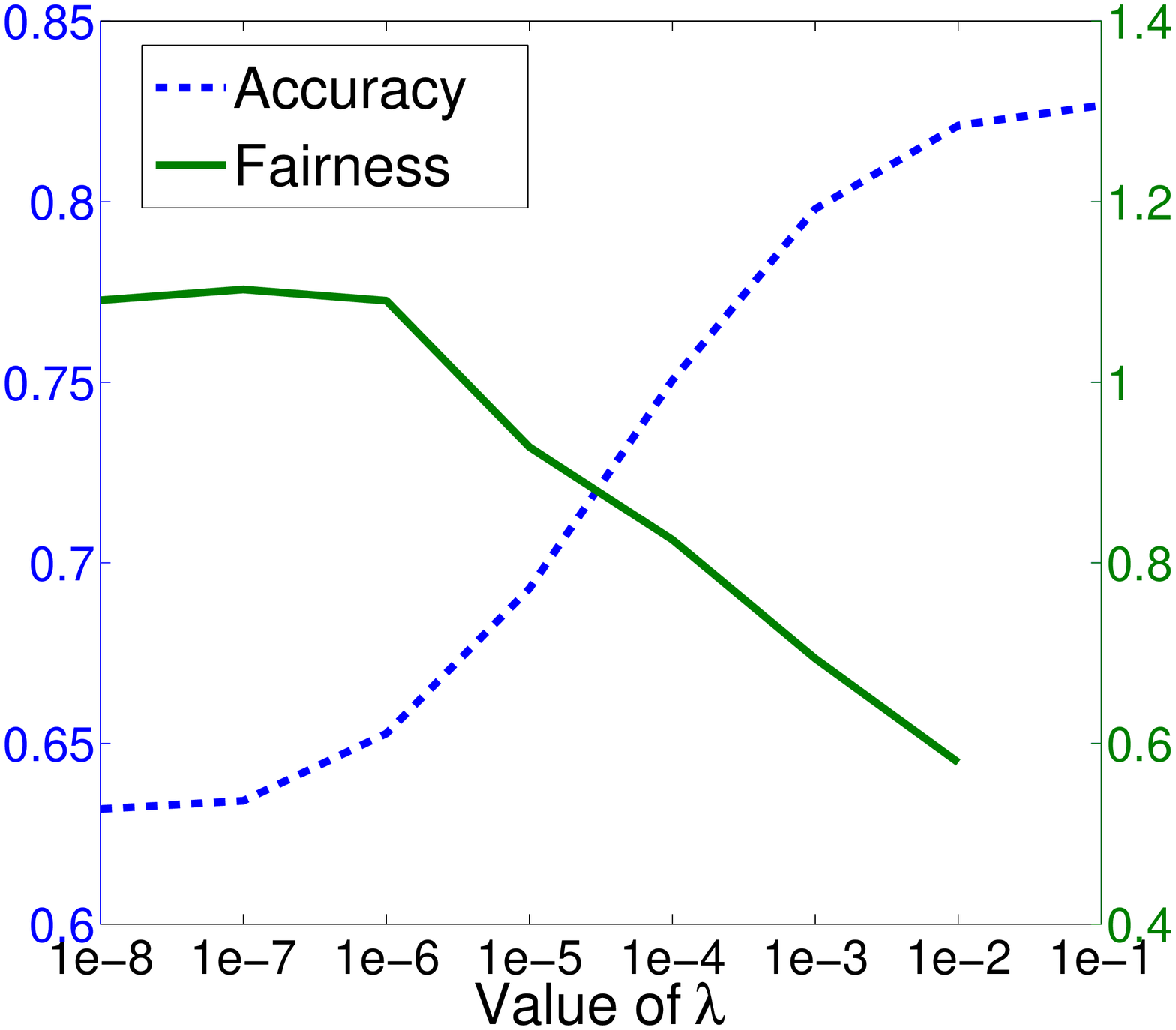}
\vspace{-15pt}
\caption{Task 1}
\end{subfigure}
\begin{subfigure}{0.23\textwidth}
\includegraphics[width=\textwidth]{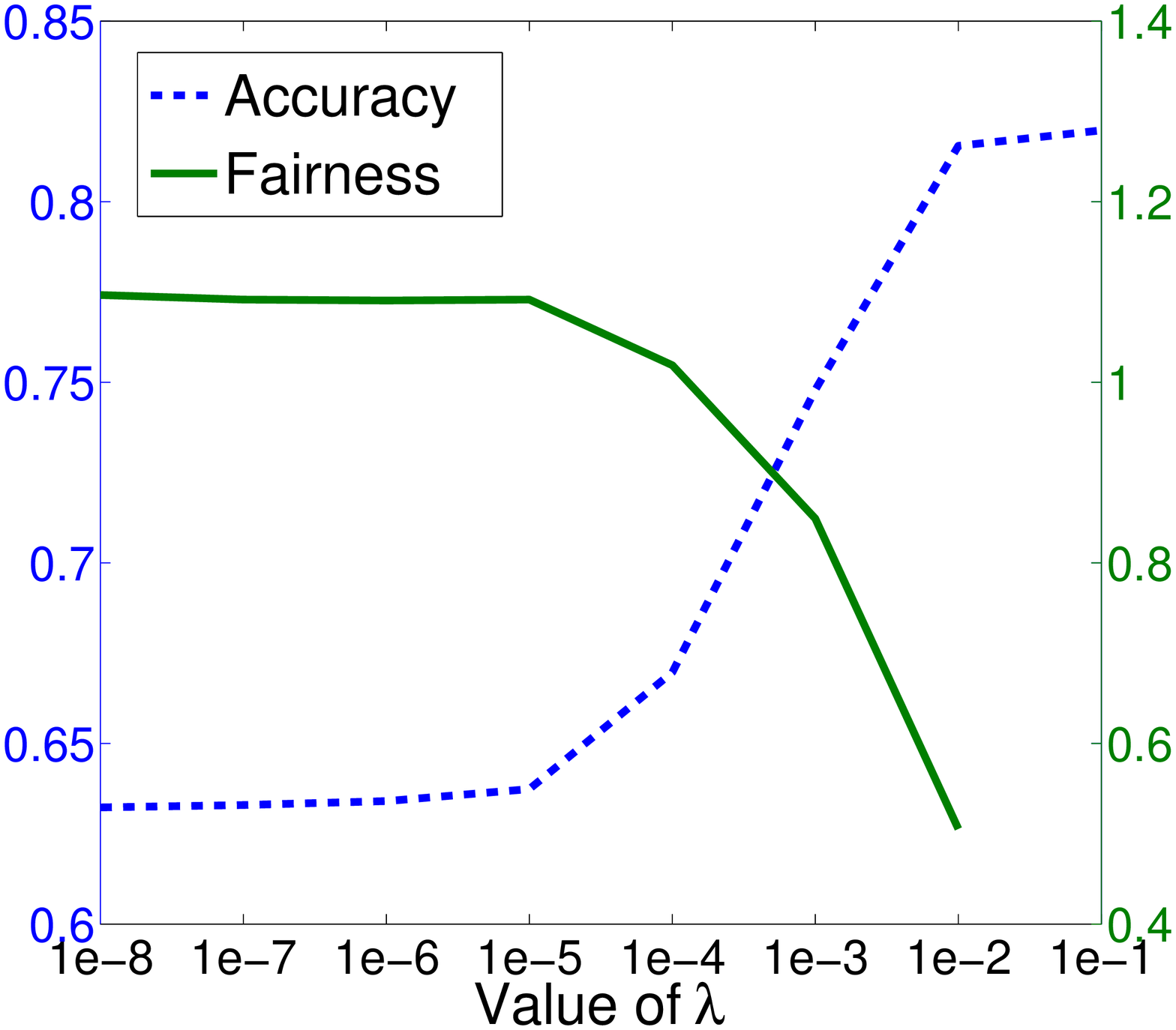}
\vspace{-15pt}
\caption{Task 2}
\end{subfigure}
\caption{M.T. Performance under Setting 1.} 
\label{mtltwotasks}
\end{figure}

Next, we examined the performance of multi-task algorithm 
(\ref{mtlalgorithm}). The source code provides 
multiple options of the output kernel learning modalities, 
and we only presented results based on the Frobenius modality 
(as similar trends were observed using other modalities).  
The performance of each task in setting 1 is shown 
in Figure \ref{mtltwotasks}, where we gradually increased 
$\lambda$ to enforce stronger task relation. 

\begin{figure}[t!] 
\centering 
\begin{subfigure}{0.23\textwidth}
\centering
\includegraphics[width=\textwidth]{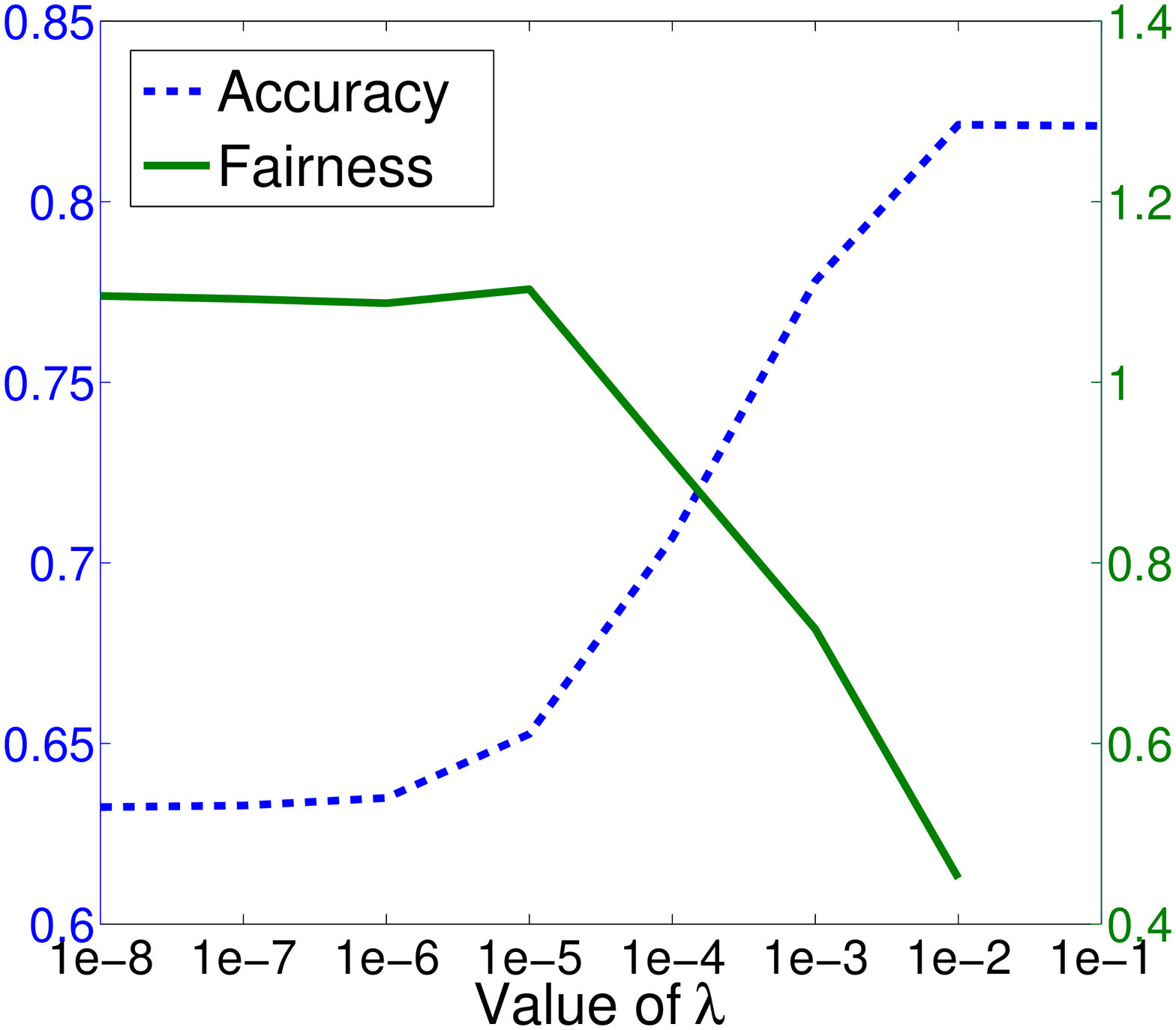}
\vspace{-15pt}
\caption{Setting 1} 
\end{subfigure}
\begin{subfigure}{0.23\textwidth}
\centering
\includegraphics[width=\textwidth]{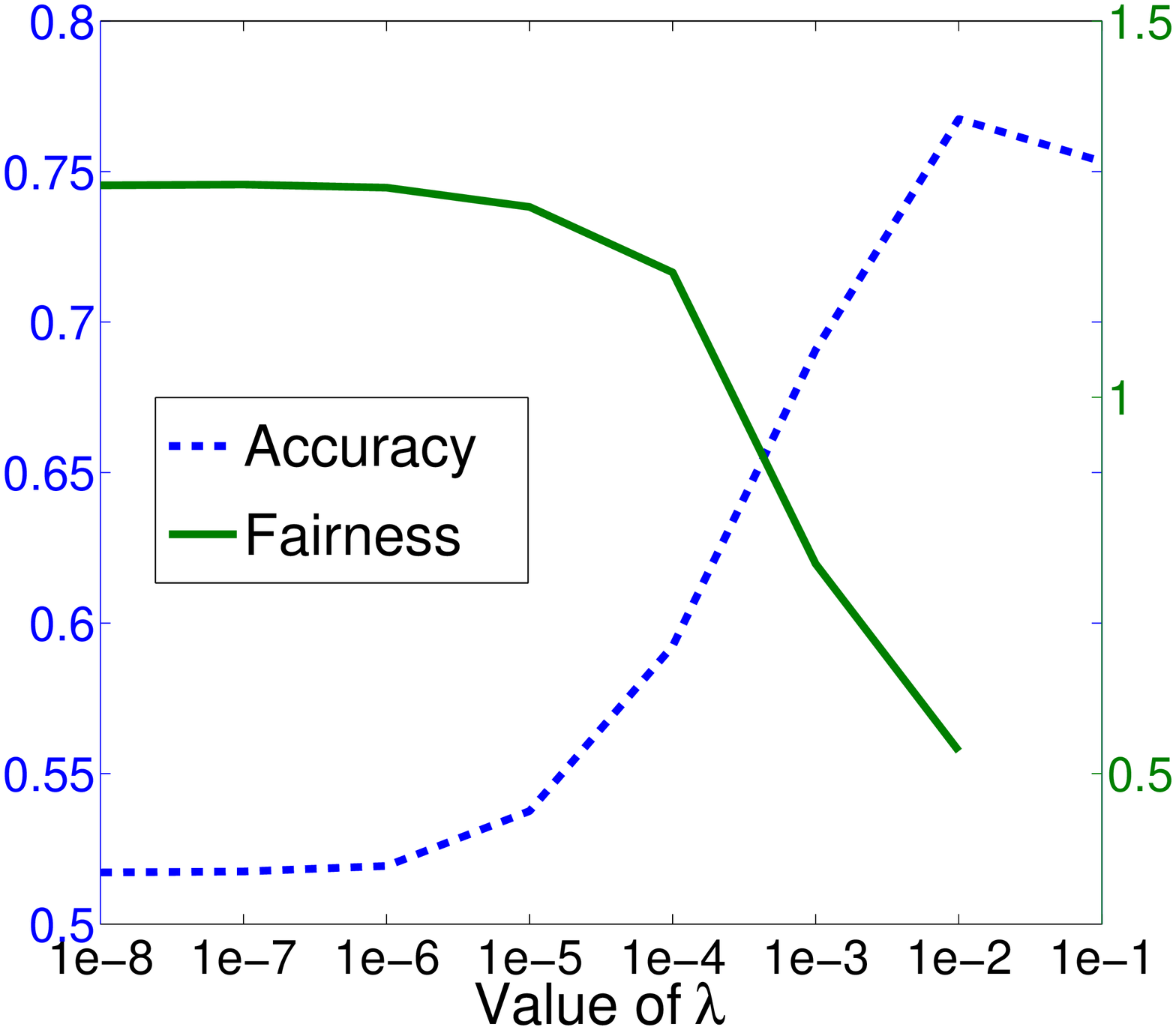}
\vspace{-15pt}
\caption{Setting 2}
\end{subfigure}
\begin{subfigure}{0.23\textwidth}
\centering
\includegraphics[width=\textwidth]{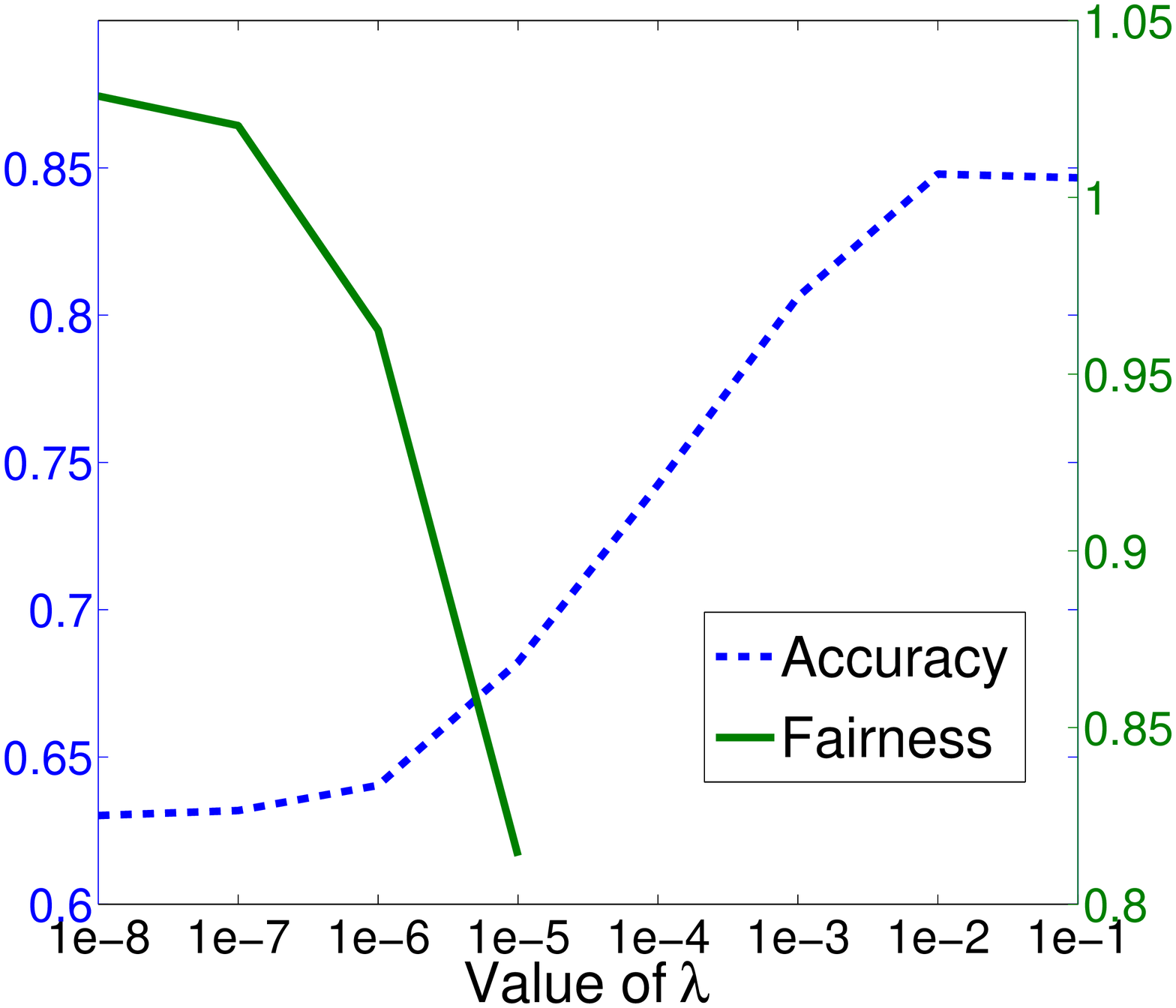}
\vspace{-15pt}
\caption{Setting 3}
\end{subfigure}
\begin{subfigure}{0.23\textwidth}
\centering
\includegraphics[width=\textwidth]{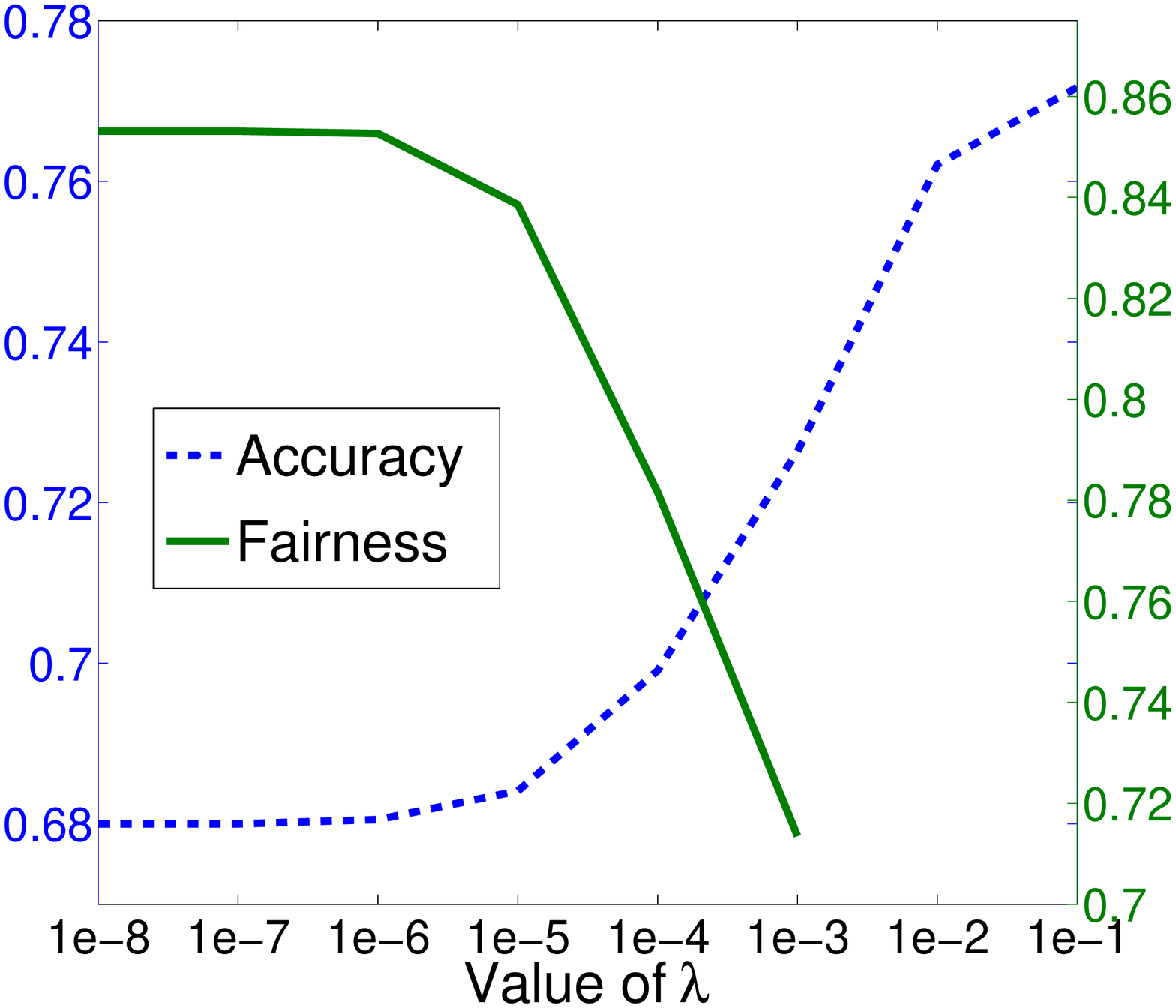}
\vspace{-15pt}
\caption{Setting 4}
\end{subfigure}
\caption{M.T. Performance under all Four Settings. Results 
are averaged over two tasks.} 
\label{mtlallset}
\end{figure}

Discriminatory transfer can be seen in each task, 
i.e. as $\lambda$ increases, both tasks have improved 
prediction accuracies but lowered fairness. 
Similar phenomena are observed in other settings, as 
shown in Figure \ref{mtlallset}. 

\section{Discussion}

The presented study shows standard transfer learning 
can improve prediction accuracy of target tasks at the 
cost of lowering their prediction fairness. How to explain 
this phenomenon in principle, and how to minimize fairness 
damage while retaining the traditional accuracy improvement
gained by transfer learning remain open questions. 

Zemel et al \cite{zemel2013learning} had a very 
interesting result suggesting fairness may be transferrable: 
they showed a fair feature representation learned for one 
task can be used in another task to improve the latter's 
prediction fairness. We studied a fundamentally different 
problem, and showed  discrimination can be induced from 
the transfer process, even when the pre-learned hypothesis is fair. 
Besides, their transfer setting is different from ours: 
they focused on feature learning and treated different 
label sets of the same population as different tasks, 
while ours focused on predictive learning and treated 
the same label set of different populations as different 
tasks -- a setting of domain adaptation \cite{pan2010survey}. 

The present study is motivated by negative transfer proposed 
by Rosenstein et al 
in \cite{rosenstein2005transfer}; 
they showed learning tasks jointly may not improve their 
prediction accuracies if the tasks are not as similar as presumed. 
We had not come up with a similarly concrete hypothesis for 
discriminatory transfer, however, and  only conjectured 
transfer process may rule out fair hypotheses when biasing 
target task learning.

\begin{acks}
  The authors would like to thank anonymous reviewers 
  for their valuable comments and suggestions. 
\end{acks}

\bibliographystyle{ACM-Reference-Format}
\bibliography{fairness,multitask,dataset} 

\end{document}